\documentclass{aa}
\usepackage{epsfig}
\usepackage{rotating}
\usepackage{natbib}
\usepackage{graphicx}

\newcommand{\chan}{{\sl Chandra}}

\newcommand{\xmm}{{\sl XMM-Newton}}

\newcommand{\hst}{{\sl HST}}
\newcommand{\vlt}{{\sl VLT}}

\newcommand{\spitzer}{{\sl Spitzer}}

\newcommand{\naco}{{\sl NACO}}
\newcommand{\nicmos}{{\sl NICMOS}}
\newcommand{\nacon}{{\sl NAos COnica}}

\newcommand{\tmass}{{\sl 2MASS}}
\newcommand{\gsc}{{\sl GSC-2}}

\newcommand{\eclipse}{{\sl eclipse}}

\begin{document}

 \title{\vlt/\naco\ observations of the High-Magnetic field radio
 pulsar PSR J1119-6127\thanks{Based on observations collected at the
 European Southern Observatory, Paranal, Chile under programme ID
 076.D-0613(A) } }

 \author{R. P. Mignani\inst{1}
 \and 	
 R. Perna\inst{2}
\and
 N. Rea\inst{3,4}
\and
 G.L. Israel \inst{5}
\and
 S. Mereghetti\inst{6}
\and 
G. Lo Curto\inst{7}
}

   \offprints{R. P. Mignani}

   \institute{University College London, Mullard Space Science Laboratory, Holmbury St. Mary, Dorking, Surrey, RH5 6NT, UK\\
              \email{rm2@mssl.ucl.ac.uk}
\and
JILA and Department of Astrophysical and Planetary Sciences, University of Colorado, 440 UCB, Boulder, 80309, USA
\and
 SRON Netherlands Institute for Space Research, Sorbonnelaan 2, 3584 CA Utrecht, The Netherlands
\and
University of Sydney, School of Physics A29, NSW 2006, Australia
\and
  INAF Astronomical Observatory of Rome, Via di Frascati 33, 00040, Monte Porzio, Italy 
\and
Istitituto di Astrofisica Spaziale, Via Bassini 15, 20133, Milan, Italy 
\and
European Southern Observatory, Alonso de Cordova 3107, Vitacura, Santiago, Casilla 19001 Santiago 19, Chile
}

     \date{Received ...; accepted ...}

     \abstract {Recent radio  observations have unveiled the existence
       of a  number of radio  pulsars with spin-down  derived magnetic
       fields in the  magnetar range. However, their observational
       properties  appears  to  be  more similar  to  classical  radio
       pulsars than  to magnetars.}{ To  shed light on this  puzzle we
       first have to determine  whether the spin-down derived magnetic
       field values for these  radio pulsars are indeed representative
       of  the actual  neutron  star  magnetic field  or  if they  are
       polluted,  e.g.  by  the effects  of a  torque from  a fallback
       disk.}{To investigate this  possibility, we have performed deep
       IR ($J,H,K_s$ bands) observations of one of these high magnetic
       field  radio pulsars  (PSR J1119--6127)  with the  ESO  \vlt\ to
       search for  IR emission which  can be associated with  a 
       disk.}{No IR emission is detected from the pulsar position down
       to $J  \sim$24, $H \sim$23 and $K_s  \sim$22.}{By comparing our
       flux upper limits with the predictions of fallback disk models,
       we  have found  that  we can  only  exclude the  presence of  a
       disk with accretion rate $\dot{M}\ga 3\times 10^{16}$ g s$^{-1}$.
       This lower limit cannot rule  out the presence of a substantial
       disk  torque   on  the  pulsar,   which  would  then   lead  to
       overestimate the value of  the magnetic field inferred from $P$ and $\dot{P}$.
       We have also compared the upper limit  on the IR  luminosity of PSR
       J1119--6127    with   the  IR   luminosities    of
       rotation-powered pulsars  and magnetars.  We  found that, while
       magnetars  are intrinsically  more efficient  IR  emitters than
       rotation-powered  pulsars,  possibly  because of  their  higher
       magnetic field,  the relatively  low IR emission  efficiency of
       PSR J1119--6127 suggests that it is more similar to the latters than to the
       former. 
}


\titlerunning{\vlt\ observations of PSR J1119--6127}

   \maketitle

\section{Introduction} 

High-energy  observations  performed  in  the last  two  decades  have
unveiled the existence of a few unusual classes of neutron stars (NSs;
see Popov 2007 for a  review).  The Anomalous X-Ray Pulsars (AXPs) and
the Soft Gamma-Ray Repeaters (SGRs)  are among them the most peculiar
objects (see Woods \& Thompson  2006 for a recent review). At variance
with the  majority of the  NSs known so  far, i.e. the  radio pulsars,
they are typically radio quiet but show X-ray pulsations at periods of
a few seconds. Furthermore, the X-ray luminosity of both SGRs and AXPs
largely  exceeds  their rotational  energy  ($L_X  \approx 100  \times
\dot{E}$),  while the rotational  energy of  radio pulsars  can easily
account  for their X-ray  emission ($L_X  \sim 0.001  \times \dot{E}$;
Becker \& Tr\"umper  1997; Possenti et al.  2002).   The properties of
AXPs  and  SGRs  are  well  explained  by  the  magnetar  model  which
interpretes  these objects  as  isolated neutron  stars with  magnetic
fields $B \sim 10^{14-15}$~G (hence dubbed magnetars), consistent with
their  observed   spin-down  with  the  usual   vacuum  dipole  losses
\footnote{Through the  paper magnetic  fields are computed (in G) using the
relation $B=3.2\times 10^{19}  (P \: \dot P)^{1/2}$, where  $P$ is the
NS period (s), $\dot P$ is its spin-down rate (s s$^{-1}$).}.  In the magnetar model,
the X-ray  luminosity is thought to  be powered by  the magnetic field
decay,  while  radio pulsations  were  believed  to  be suppressed  by
processes such as the  photon splitting, which inhibit pair-production
cascades  in  magnetic  fields  greater than  the  ``quantum  critical
field'' $B_c=4.4\times 10^{13}$~G (Baring \& Harding 1998).

This  dichotomy  between the  two  different  pulsar  classes -  radio
pulsars with $B<  B_c$ on one side, and magnetars  with $B>B_c$ on the
other -  was shaken  by the discovery  of radio pulsars  with magnetic
fields above  $B_c$ (Camilo et  al.  2000).  Despite having  such high
magnetic  fields, although lower  than those  of the  magnetars, these
high-magnetic field  radio pulsars (HBRPs) do not  behave according to
any of  the known magnetars templates.   First of all,  {\it they are}
radio pulsars, while pulsed radio  emission has been discovered so far
only  in  the  transient   magnetar  XTE  J1810--197  (Camilo  et  al.
2006). Second, only two  HBRPs, PSR J1119--6127 (Gonzalez \& Safi-Harb
2003) and  PSR  J1718--3718  (Kaspi  \& McLaughlin  2005),  have  been
detected  in X-rays so  far, with  luminosities $L_X  \sim 10^{32-33}$
erg s$^{-1}$ almost  two orders of magnitude lower  than those of the
magnetars and smaller than their $\dot E$.  Finally, HBRPs do not show
bursting emission,  either in X-rays  or in $\gamma$-rays,  while AXPs
and SGRs  instead do.  These  differences might be  explained assuming
e.g.,  that  HBRPs are  dormant  transients,  that  their lower  X-ray
luminosities  are a  consequence of  their lower  magnetic  fields, or
simply assuming  that different  evolutionary paths or  stages account
for the different phenomenologies.

Of  course,  one  alternative  possibility is  that  the  spin-derived
magnetic field values  of the HBRPs are unreliable  because e.g., they
are overestimated by the extra torque produced by a fossil disk formed
out  of residual matter  from the  supernova explosion.   Fossil disks
around  isolated NSs have  been invoked  over the  years to  explain a
large  variey of  phenomena (e.g.  Michler  \& Dressler  1981; Lin  et
al. 1991;  Phinney \& Hansen  1993; Podsiadwolski 1993;  Chatterjee et
al.   2000; Alpar 2001;  Menou et  al. 2001;  Blackman \&  Perna 2004;
Cordes  \&  Shannon  2006), and  at  least  in  the  case of  the  AXP
4U\,0142+61,  recent  \spitzer\  observations  possibly  revealed  the
presence of one of these disks (Wang et al.  2006).  Thus, if HBRPs do
have   fossil  disks,   they  should   be  detectable   through  deep,
high-resolution  IR  observations.   Since  the  IR  luminosity  of  a
hypothetical disk  is expected to  be larger for X-ray  bright pulsars
due  to the  flux  contribution  from the  reprocessing  of the  X-ray
radiation (Perna et  al.  2000; Perna \& Hernquist  2000), the primary
candidates are obviously the HBRPs detected in X-rays.

In  this  work  we  report  on  the results  of  our  recent  deep  IR
observations  of PSR J1119--6127.   The pulsar  was discovered  in the
Parkes multi-beam survey (Camilo et al. 2000) with period $P = 407$ ms
and period  derivative $\dot P \sim  4.022 \times 10 ^{-12}$  s s $^{-1}$,
which give  a characteristic  age of $\sim  1600$ years,  a rotational
energy loss $\dot E \sim 2.3 \times 10 ^{36}$ erg s$^{-1}$, and a magnetic
field $B  \sim 4.1 \times 10^{13}$ G.  PSR J1119--6127 is also  one of the
very few pulsars with a measure  of the braking index of $2.9 \pm 0.1$ (Camilo et al. 2000).
X-ray emission  was first detected with \chan\  (Gonzalez \& Safi-Harb
2003) which also  revealed a compact  pulsar wind nebula,  while X-ray
pulsations were discovered with \xmm\ (Gonzalez et al.  2005).  \\ The
structure of the paper is  as follows: IR observations and results are
described  in \S2,  while comparisons  with  disk models  and with  IR
observations  of other  isolated NSs  are  discussed in  \S3 and  \S4,
respectively.

\begin{table*}
\begin{center}
  \caption{Summary of  the \naco\ $J,H,K$-band observations of  the PSR J1119-6127
field with the number of exposure sequences, the total number of exposures per filter, the DIT and NDIT, the average seeing and airmass. }
\begin{tabular}{lccccccc} \hline
yyyy.mm.dd     & Filter & N & $N_{exp}$ & DIT (s) & NDIT & Seeing (``) & Airmass	\\ \hline
2006.01.25     & $K_s$  & 1 & 8  & 20      & 16     &0.76 & 1.26  \\
2006.02.23     & $K_s$  & 2 & 30 & 55      & 3      &0.66 & 1.30  \\
2006.02.24     & $H$    & 2 & 30 & 55      & 3      &0.61 & 1.29  \\
               & $J$    & 2 & 30 & 55      & 3      &0.88 & 1.27  \\
2006.02.28     & $J$    & 1 & 15 & 55      & 3      &0.59 & 1.32  \\
\hline
\end{tabular}
\label{nacodatasummary}
\end{center}
\end{table*}

\section{IR Observations}

\subsection{Observations Description}

IR observations of PSR  J1119--6127 have been performed in Service
Mode on  January 25th, February 23rd,  24th and 28th  2006   with \nacon\  (\naco), an adaptive  optics (AO)  imager and
spectrometer mounted at  the fourth Unit Telescope (UT4)  of the \vlt.
In order  to provide the  best combination between  angular resolution
and sensitivity, \naco\  has been operated with the  S27 camera with a
corresponding field of  view of $28''\times28''$ and a  pixel scale of
0\farcs027.  As  a reference  for the AO  correction we have  used the
\gsc\ star  S111230317098 ($V=13.7$), located 29\farcs5  away from our
target.   Unfortunately,  no  suitable  reference star  was  available
within  the  small  \naco\ S27  field  of  view,  which makes  our  AO
correction not optimal and  more sensitive on small scale fluctuations
of the  atmospheric conditions.   The Visual ($VIS$)  dichroic element
and   wavefront  sensor   ($4500-10000  \:   \AA$)  have   been  used.
Observations have been performed  in the ESO Johnson $J (\lambda=12650
\: \AA  ; \Delta \lambda=  2500 \: \AA)$,  $H (\lambda=16600 \:  \AA ;
\Delta \lambda= 3300 \: \AA)$  and $K_s (\lambda=21800 \: \AA ; \Delta
\lambda= 3500 \: \AA)$ filters.

To  allow for  subtraction of  the  variable IR  sky background,  each
observation has been split in two sequences of short randomly dithered
exposures with  Detector Integration Times (DIT)  of 20 and  55 s, and
NDIT repetitions along each point  of the dithering pattern (see Table
\ref{nacodatasummary}).  This  yields a total net  integration time of
about 2500 s per band,  per exposure sequence.  For each exposure, the
instrument readout mode has been selected according to the used DIT in
order to minimize the read out noise.  Owing to the expected faintness
of the  target, the DIT/NDIT  combination has been modified  after the
$K_s$  band observation  of  the first  night  to allow  for a  better
signal-to-noise in the single exposures  and to allow for a better hot
pixels rejection.

For all our observations, the seeing conditions were on average below
$0\farcs8$ and the airmass was better than 1.3, allowing for a better
yield of the \naco\ adaptive optics.  Sky conditions were photometric
in both nights.  Night (twilight flat fields) and day time calibration
frames (darks, lamp flat fields) have been taken daily as part of the
\naco\ calibration plan.  Standard stars from the Persson et al.
(1998) fields have been observed at the beginning of all nights for
photometric calibration. As we expect the photometry errors to be
dominated by the target's counts statistic rather than by the accuracy
of the photometric calibration, we have not acquired photometric
standard star fields prior to each exposure sequence.

\subsection{Data Reduction and Analysis}

The    data   have    been    processed   using    the   ESO    \naco\
pipeline\footnote{http://www.eso.org/observing/dfo/quality/NACO/}
and the science images reduced  with the produced master dark and flat
field  frames.  For  each band,  and  for each  night, single  reduced
science exposures  have been combined  to produce cosmic-ray  free and
sky-subtracted images.   The photometric calibration  pipeline yielded
average zero points  of $23.03 \pm 0.02$ and  $23.08 \pm 0.03$ ($K_s$)
for  January 25th and  February 23rd,  respectively, $24.08  \pm 0.04$
($J$)  and $23.94 \pm  0.04$ ($H$)  for February  24th, and  $24.1 \pm
0.05$  ($J$)   for  February  28th.    The  data  have   been  reduced
independently    using   procedures    run    under   the    \eclipse\
package\footnote{http://www.eso.org/projects/aot/eclipse/}     yielding
qualitatively similar data.

As a  reference for the  position of PSR  J1119--6127 we have  used its
radio  coordinates $\alpha  (J2000)$=11$^h$ 19$^m$  14.30$^s$, $\delta
(J2000)$=  -61$^\circ$  27'  49\farcs5,  which  have  an  accuracy  of
0\farcs2 (Camilo  et al.  2000).   The astrometry on the  \naco\ image
have  been computed using  as a  reference 7  stars selected  from the
\tmass\  catalogue. The  pixel  coordinates of  these  stars (all  non
saturated and evenly  distributed in the field) have  been measured by
gaussian fitting their intensity  profiles using the specific function
of    the   GAIA    (Graphical   Astronomy    and    Image   Analysis)
tool\footnote{star-www.dur.ac.uk/~pdraper/gaia/gaia.html}   while  the
fit to the $\alpha$,$\delta$  reference frame has been performed using
the                          Starlink                          package
ASTROM\footnote{http://star-www.rl.ac.uk/Software/software.htm}.    The
rms of the  astrometric solution turned out to  be $\approx$ 0\farcs09
per coordinate.  After accounting for the 0\farcs2 average astrometric
accuracy of \tmass\
\footnote{http://spider.ipac.caltech.edu/staff/hlm/2mass/},
the overall uncertainty to be attached to the position of our target
is finally 0\farcs3.

\begin{figure*}
\centering 
\includegraphics[width=8.0cm,width=16cm,angle=0,clip=]{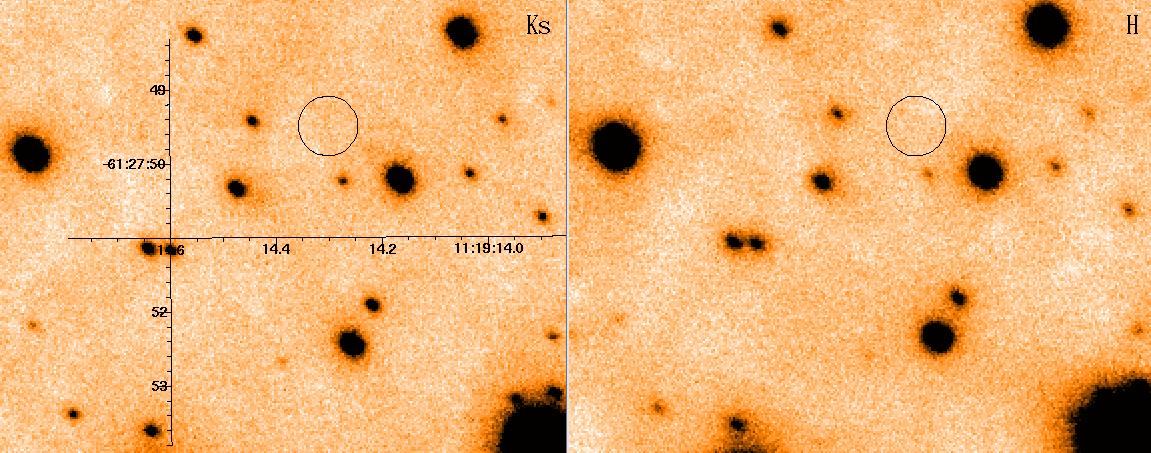}
\caption{$6'' \times  6''$ sections of  the \vlt/\naco\ $K_s$  and $H$
band images of the PSR J1119-6127 field. North to the top, East to the
left. The effect of the worse AO correction (see \S 2.1) is recognized
from  the  asymmetric  PSF of  the  stars  in  the field.  The  circle
($0\farcs3$   radius)  corresponds  to   the  pulsar   radio  position
uncertainty  after  accounting for  the  accuracy  of our  astrometric
solution (see \S2.2).  }
\label{psr_naco}       
\end{figure*}

\subsection{Results}

Fig.~ \ref{psr_naco} shows the $K_s$ and $H$ band images of the PSR
J1119--6127 field with the computed pulsar radio position overlaid. No
potential counterpart is detected at the expected position, with the
closest object being detected $\sim 1 \sigma$ away from the edge of
the error circle. The same is true also for the $J$ band image. We
thus conclude that both the pulsar and its putative disk are
undetected in each of the three observing bands down to estimated
limiting magnitudes of $J \sim$24, $H \sim$23 and $K_s \sim$ 22.  At
the same time, no diffuse emission is recognized which can be possibly
associated with the X-ray pulsar wind nebula detected by \chan\
(Gonzalez \& Safi-Harb 2003).

\section{Discussion}

\subsection{Comparison with disk models}

We have used the derived IR flux upper limits to constrain the range
of parameters that a hypothetical fossil disk around the pulsar could
have.   If a disk were indeed present and interacting with the
pulsar magnetosphere, then, as mentioned in \S1 and detailed below,
the $B$ field inferred from $P$ and $\dot{P}$ could be largely
overestimated.  The torque exterted by a disk on the star
magnetosphere can be written as (e.g. Menou et al. 2001) $\dot{J}_{\rm
disk}=I\dot{\Omega}\sim -2\dot{M}R^2_{\rm in}\Omega$, where $\dot{M}$
is the disk accretion rate, $R_{\rm in}$ is the disk inner radius, and
$\Omega=2\pi/P$ is the angular frequency of the pulsar.  The fact that
PSR J1119--6127 is detected in radio implies that $R_{\rm in}$ cannot
be smaller than the light cylinder radius $R_{\rm lc}=c/\Omega$
(e.g. Illarionov \& Sunyaev 1975).
On the other hand, if the inner radius of
the disk were outside of the light cylinder, where the magnetic
field lines are open, no efficient torque could operate. Therefore,
in the following analysis we consider only the case
$R_{\rm in}= R_{\rm  lc}$, which yields a torque $\dot{J}_{\rm
disk}=-2\dot{M}c^2/\Omega$ or, equivalently,   
an energy loss   (in   modulus)   $\dot{E}_{\rm   disk}\sim   2\dot{M}   c^2   =
2\;10^{37}\;\dot{M}/(10^{16}{\rm g  \: s^{-1}})$ erg  s$^{-1}$.  Under
these conditions, the total energy  loss of the pulsar, accounting for
both  the  dipole   and  the  disk  torque  components,   is  given  by
$\dot{E}=\dot{E}_{\rm   dip}  +  \dot{E}_{\rm   disk}\sim  B^2\Omega^4
R^6/6c^3+2\dot{M} c^2$, where $R$ is the radius of the star.  Clearly,
if $\dot{E}_{\rm  disk}\ga \dot{E}_{\rm dip}$, the value  of $B$
that   is   inferred   from   $P$  and   $\dot{P}$,   i.e.    assuming
$\dot{E}=\dot{E}_{\rm dip}$,  could be largely  overestimated.  In the
case  of  PSR J1119--6127,  a  fallback disk with accretion  rate  of  the order  of
10$^{15}$ g s$^{-1}$  could account for the entire  energy loss of the
pulsar ($\dot E \sim 2.3  \times 10 ^{36}$ erg s$^{-1}$), even without
the contribution of dipole losses, which would be the case for a very low magnetic field. We
thus take  $\dot{M} \sim  10^{15}$ g s$^{-1}$  to be the  accretion rate
corresponding  to the  maximum  torque that could be produced by a  hypothetical
disk.

We have simulated the disk IR spectrum using the disk model developed
by Perna et al. (2000), which takes into account the contribution to
the disk IR emission due to both viscous dissipation and reprocessing
of the X-ray radiation from the pulsar.  The spectra have been
renormalized for the distance $d$ to PSR J1119--6127.  Camilo et al.
(2000) reported $d=2.4-8$ kpc, while Gonzalez \& Safi-Harb (2003),
based on the measured extinction per unit distance in the pulsar
direction, estimated $d=5.4-12.6$ kpc.  Most likely, the pulsar is not
further than 8 kpc, according to its location with respect the Carina
spiral arm (Camilo et al.  2000). In the following we report our
results as a function of D$_6$=$d$/(6 kpc).  
Fig.~2 shows the modelled
disk IR spectra computed for two different values of the disk
accretion rate $\dot{M}$ compared to the observed IR flux upper
limits.  In particular, we show the predicted flux corresponding
to the maximum value of $\dot{M}$ that would make the disk emission
compatible with the current limits, as well as the flux corresponding
to the maximum value of $\dot{M}$ compatible with the spin down rate
of the pulsar.

\begin{figure}[t]
\centering
\includegraphics[bb=5 120 610 725,width=8.0cm,angle=0,clip=]{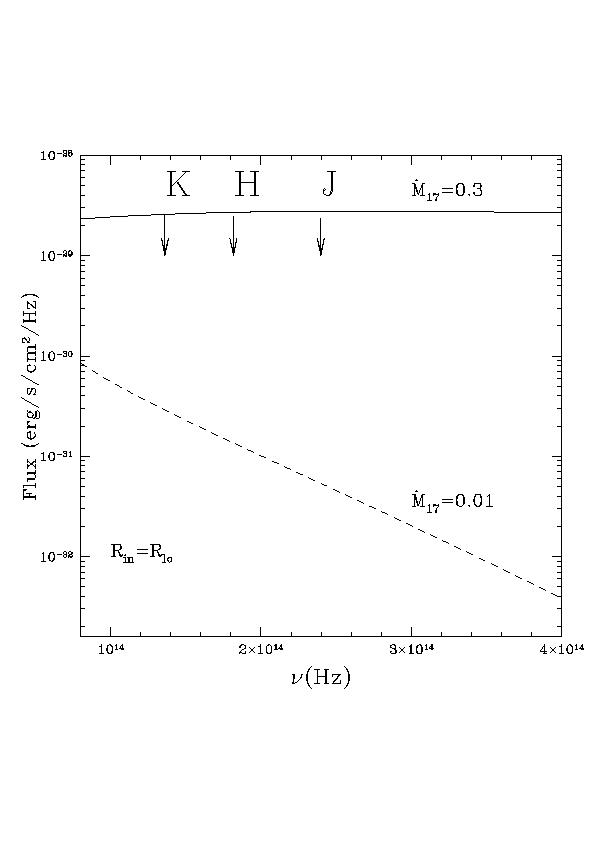}
\caption{Expected IR spectrum of a fallback disk of inner radius
$R_{\rm in}=R_{\rm lc}$  and two values of the disk accretion rate
$\dot{M}$: {\em solid line:} maximum value of $\dot{M}$ compatible
with the IR limits; {\em dashed line:} maximum value of $\dot{M}$
compatible with the pulsar's spin down rate.  The IR flux upper
limits have been corrected for the interstellar extinction applying
the relations of Fitzpatrick (1999) for $A_V$= 5, as derived from the
X-ray absorption $N_H = 9^{+5}_{-3} \times 10^{21}$ cm$^{-2}$
(Gonzalez \& Safi-Harb 2003) and the relation $A_V = N_{H}/1.79 \times
10^{21}$ atoms cm$^{-2}$ mag $^{-1}$ (Predhel \& Schmitt 1995).  }
\label{mdot}       
\end{figure}

We find that  our limits only rule out  disks with $\dot{M}\ga 3\times
10^{16}$ D$_{6}^{2}$  g s$^{-1}$,   i.e. well  above the  value of
$10^{15}$ g s$^{-1}$ corresponding to the case of maximum allowed disk
torque.  Therefore,  we cannot exclude with  the current observations
that the magnetic  field derived from the pulsar  spin-down, under the
assumption  of a purely  vacuum dipole  energy loss,  be overestimated
 due to pollution by a disk torque.   We note that, given the
low  X-ray  luminosity  of  PSR J1119--6127,  $L_{{\rm0.5-10  keV}}  =
5.5^{+10}_{-3.3} \times 10^{32} \times D^2_{6}$ erg s$^{-1}$ (Gonzalez
\& Safi-Harb  2003), the contribution to  the disk IR  emission due to
the X-rays reprocessing is so low that it becomes comparable with that
due to  viscous dissipation in  the disk only for  $\dot{M}\la 5\times
10^{15}$ D$_{6}^{2}$ g s$^{-1}$.  For this reason, the derived IR flux
upper limits  for PSR J1119--6127 are  less stringent in  ruling out a
fallback disk  at the  light cylinder with  respect to  similarly deep
upper limits obtained for the  AXPs which, instead, have a much higher
X-ray luminosity. Indeed,  an X-ray luminosity higher by  a factor 100
would  raise the disk  IR emission  much closer  to our  present upper
limits.

We have seeked for other evidence which might indirectly unveil the existence of a fallback disk and the effect of its torque on the pulsar's spin down. In principle, a  torque from a fallback disk should leave a signature  in the
pulsar timing  by increasing  the level of  the timing noise.   In the
case of PSR J1119--6127, the level  of the radio timing noise does not
show any clearly  anomalous excess which might be  associated with the
effect of  an acting disk  torque, and it seems  apparently consistent
with  the  level  expected  for  its high  $\dot  P$  (Arzoumanian  et
al. 1994).   However, we note that  the magnitude of  the effect would
depend  on the actual  value of  the disk  torque, which  is obviously
unknown, and it might be confused with the underlying timing noise.
In any case, the timing analysis can in no way rule out that the pulsar's
spin down might have been affected by a disk torque in the past.

\subsection{Comparison with other NSs}

\begin{table*}[t]
\begin{center}
  \caption{Summary of the  IR fluxes measurements for  all types of isolated NSs with an identified IR  counterpart i.e.  rotation-powered pulsars
    (rows  1-5)  and magnetars  (rows  6-11).   The  columns give  the
    observed  $J,  H,  K,   K_s$  magnitudes  (an  hyphen  stands  for
    non-detection,  values  in italics  have  been extrapolated),  the
    distance,  and the  interstellar extinction  $A_V$  either derived
    from existing  optical measurements (O) or from  the $N_H$ derived
    from the  fits to the X-ray  spectra (X) by using  the relation of
    Predehl \&  Schmitt (1995). $K$-band  flux values in  italics have
    been  derived  from the  extrapolation  of  the  $J$ and  $H$-band
    fluxes. }
\begin{tabular}{l|cccc|l|l|l} \hline
NS Name            & $J$         & $H$         & $K$         & $K_s$       &$d{\rm(kpc)}$        & $A_V$   &  Ref.\\ \hline
Crab                &14.8$\pm$0.05&14.3$\pm$0.05&13.8$\pm$0.05& -           &1.730$\pm$0.28           &1.62 (O)        &1,2,3\\
PSR B1509--58        &   -         &20.6$\pm$0.20& -           &19.4$\pm$0.1 &4.181$\pm$0.60           &4.8 (O)        &4,2,5\\
Vela                &22.7$\pm$0.10&22.0$\pm$0.16&{\it 21.3$\pm$0.4}& -      &0.294$^{+0.019}_{-0.017}$&0.20 (O)        &6,7,8\\
PSR B0656+14*       &24.4$\pm$0.10&23.2$\pm$0.08&22.6$\pm$0.13& -           &0.288$^{+0.033}_{-0.027}$&0.09$\pm$0.06(O)&9,10,11\\
Geminga*            &25.1$\pm$0.10&24.3$\pm$0.10&{\it 23.4$\pm$0.4}& -      &0.157$^{+0.059}_{-0.034}$&0.12$\pm$0.09(O)&9,12,13\\ \hline
4U 0142+61**        & -	       &-            &19.7$\pm$0.05   &20.1$\pm$0.08&$\ge$5                   &5.1  (X)        &14,15\\
1E 1048.1--5937      &21.7$\pm$0.30&20.8$\pm$0.30&-            &21.3$\pm$0.30&3$\pm$1                  &6.1  (X)        &16,17,15\\
1RXS J170849--400910$^{\rm x}$   &20.9$\pm$0.10&18.6$\pm$0.10&-            &18.3$\pm$0.10&5                        &7.8  (X)        &18,15\\
XTE J1810--197       & -           &22.0$\pm$0.10&-            &20.8$\pm$0.10&4$\pm$1$^+$              &5.1  (X)       &19,20,15\\
1E 2259+586         & -	       &-	     &-               &21.7$\pm$0.20&3.0$\pm$0.5              &5.7$\pm$0.1 (O) &21,22,23\\ \hline
SGR 1806--20	    & -	       & -	     &-               &20.1$\pm$0.14&$15.1\pm1.6$             &29$\pm$2     (O) &24,25,26\\ \hline
\end{tabular}
\label{IRsummary}
\end{center}

(1) Sollerman  (2003); 
(2) radio dispersion measure, Cordes \& Lazio (2002); 
(3) Sollerman et al (2000);
(4) Kaplan \& Moon (2006);
(5) Lortet et al. (1987);
(6) Shibanov   et al. (2003); 
(7) radio parallax, Dodson et al. (2003);
(8) Mignani et al. (2003); 
(9) Koptsevich et al. (2001); 
(10) radio parallax, Brisken et al. (2003); 
(11) Pavlov  et al. (1997); 
(12) optical parallax, Caraveo et al. (1996); 
(13) Kargaltsev et al. (2005); 
(14) Hulleman et al. (2004); 
(15) present work;
(16) Wang \& Chakrabarty (2002); 
(17) Gaensler et al. (2005);
(18) Israel et al. (2003); 
(19) Israel et al. (2004);
(20) Rea et al. (2004);
(21) Hulleman et al. (2001); 
(22) Kothes et al. (2002);
(23) Woods et al. (2004);
(24) Israel et al. (2005);
(25) Mc Clure-Griffiths \& Gaensler (2005);
(26) Eikenberry et al. (2004)
\\ \\
$^*$    magnitudes refer to the \hst/\nicmos\ filters 110W, 160W, 187W, which overlap the $J$, $H$ and $K$ passbands \\
$^{**}$   $K$ and $K_s$ magnitudes have been taken at different epochs \\
$^{\rm x}$ IR counterpart still to be confirmed (Safi-Harb \&  West  2005; Durant \& van~Kerkwijk 2006; Rea et al.~2007a) \\
$^+$ revised downward to 2.5 kpc (Gotthelf \& Halpern 2005)  
\end{table*}

For the estimated ranges of distance and $N_H$ (see previous section),
our  flux  upper  limits  yield  for PSR  J1119--6127  a  $K$-band  IR
luminosity  $L_{K} \le 6.6^{+11}_{-4}  \times 10^{30}  \times D^2_{6}$
erg  s$^{-1}$.   We  have  compared  this upper  limit  with  the  IR
luminosities  of different classes  of isolated  NSs.  We  caveat here
that the nature  of the IR emission may be  different across the whole
sample.  For instance, in the case of rotation-powered pulsars, the IR
emission is thought  to be produced in the  NS magnetosphere, as shown
by their power-law  spectra (e.g. Shibanov et al.  2003), while in the
case of  the magnetars  it might be  produced by an  X-ray irradiated,
(though passive),  fallback disk, by  the magnetic field decay,  or by
curvature radiation in the magnetars' coronae (Beloborodov \& Thomspon
2007).  Nevertheless, comparing the IR properties of different classes
of  isolated  NSs can  still  be  useful  to unveil  similarities  and
diversities  which  can  be  indeed  ascribed  to  different  emission
processes and  thus be used  to disantangle, e.g.   magnetospheric and
disk emitters.

Table \ref{IRsummary} summarizes the  IR flux measurements for all the
isolated NSs  with an  IR counterpart, i.e.   rotation-powered pulsars
and  magnetars.   In  order  to  make  a  consistent  comparison  with
rotation-powered  pulsars,  which  are  persistent emitters,  for  the
magnetars we  have selected only  IR flux measurements taken  when the
X-ray source was as close as possible to quiescence.  We have include
in our compilation also the  AXP 1RXS J170849--400910, although its IR
identification  has not been  confirmed yet  (Safi-Harb \&  West 2005;
Durant  \& van~Kerkwijk  2006; Rea  et  al.~2007a), hence  we did  not
consider it in the  following analysis. The proposed identification of
1E   1841--045  (Wachter  et   al.   2004)   has  been   discarded  by
high-resolution  IR observations  (Durant 2005).   No IR  emission has
been detected so  far from the X-ray Dim  Isolated NSs (XDINSs; Mignani
et al.   2007; Lo Curto et  al. 2007; Rea  et al. 2007b) and  from any
compact central  objects (CCOs) in  SNRs (Wang, Kaplan  \& Chakrabarty
2007; Fesen, Pavlov \& Sanwal 2006).

For each object we have computed its IR luminosity either in the $K_s$
 or in  the $K$  band, as we  estimate the  error due to  the passband
 correction to be below  $\sim$ 0.1 magnitudes, i.e.  fully acceptable
 for  the  goals of  our  analysis.  For  Vela  and  Geminga, we  have
 extrapolated their $K$-band magnitudes from the IR colors.   Passband  transformations  between  different  $K$-band
 filters  have   been  neglected.    The  flux  conversion   from  the
 \hst/\nicmos\ passbands to the  Johnson's ones has been applied using
 the               \nicmos\              Units              Conversion
 tool\footnote{http://www.stsci.edu/hst/nicmos/tools/}.

For  the rotation-powered  pulsars,  distance values  have been  taken
either from  the available radio/optical parallaxes or  from the radio
dispersion          measure          (Cordes         \&          Lazio
2002)\footnote{http://rsd-www.nrl.navy.mil/7213/lazio/ne\_model/}.
For the  magnetars we have used  either the distances  of the parental
stellar  clusters or  of  the associated  supernova  remnants, or  the
distances  inferred  from the  $N_H$  (see  Table \ref{IRsummary}  and
references  therein).  For the  interstellar extinction  correction we
have applied the relations of  Fitzpatrick (1999) using as a reference
either  the  measured  $A_V$  or  the value  derived  from  the  $N_H$
recomputed from our X-ray spectral fits and the relation of Predehl \&
Schmitt  (1995).   For  the  magnetars  we  have  fitted  an  absorbed
power-law  plus a  blackbody model  (see  Tiengo et  al.~2005; Rea  et
al.~2004;  Rea et  al.~2005; Woods  et al.~2004;  Patel et  al.  2003;
Morii et al.  2003; Mereghetti et al.  2004 for further details on the
single  observations)  over the  spectral  range  2-10  keV.  All  the
$N_{H}$  values  have been  computed  assuming  solar abundances  from
Anders  \& Grevesse (1989).   Although the  reference $A_V$  have been
obtained with  different methods,  this does not  affect significantly
our estimates of  the IR luminosity, expecially in  the $K$-band where
the effects of the interstellar  extinction are lower.  The overall IR
luminosity errors take into account the measured photometric errors as
well as all  the uncertainties on the isolated NS  distance and on the
interstellar     extinction     correction,     all    reported     in
Table~\ref{IRsummary}.

\begin{figure*}[ht]
\centering 
\includegraphics[bb=10 180 440 650,width=8.0cm,angle=0,clip=]{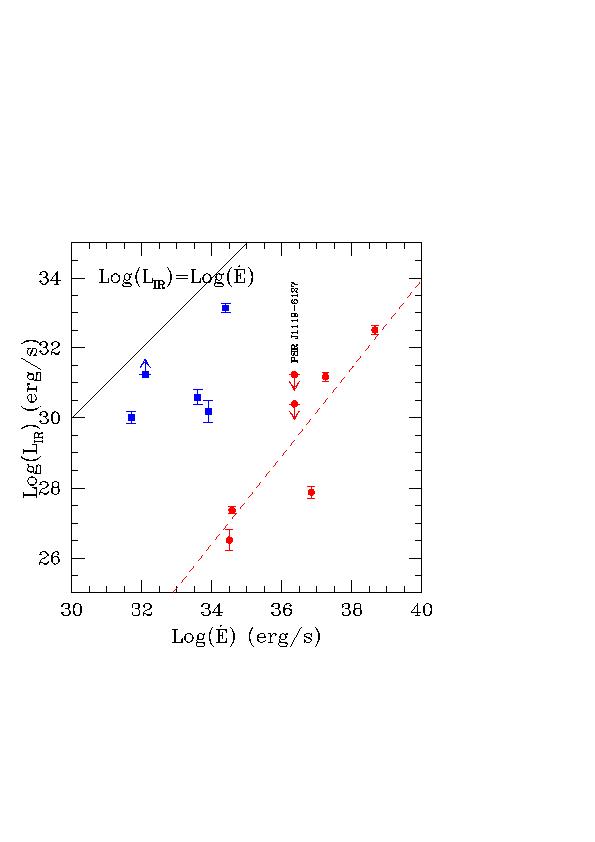}
\includegraphics[bb=10 180 440 650,width=8.0cm,angle=0,clip=]{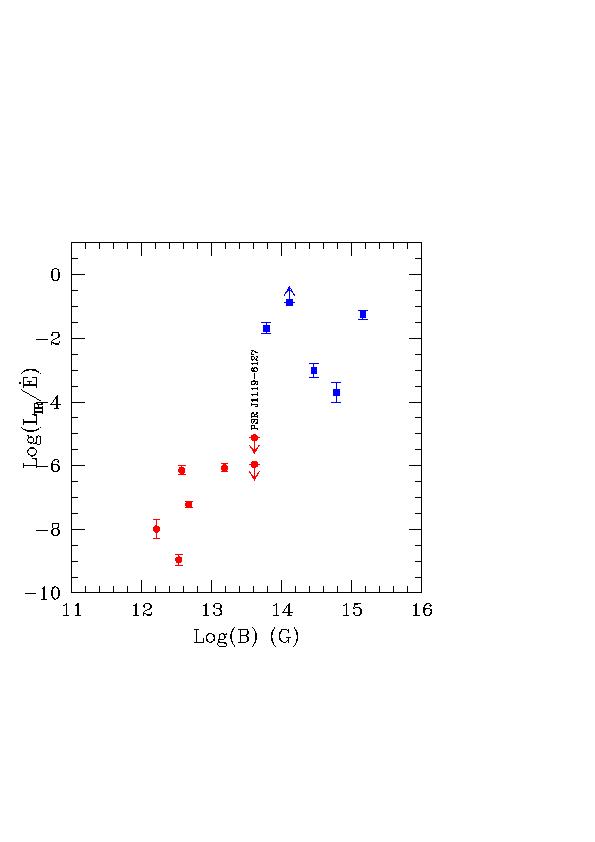}
\caption{left panel: measured $K$-band luminosities for all the isolated NSs
  listed in Table \ref{IRsummary} as  a function of the NSs rotational
  energy loss $\dot  E$.  For J1119--6127 (labelled in  the figures) we
  have plotted the  IR luminosity upper limits corresponding  to the most
  extreme  distance/absorption values.  For  4U 0142+61  we have
  plotted the IR luminosity lower limit corresponding to a lower limit on
  the source distance  of 5 kpc (Hullemann et  al.  2004).  Red filled
  circles  and blue filled  squares indicate  rotation-powered pulsars
  and magnetars, respectively.  The red dashed line corresponds to the
  linear fit  for rotation-powered pulsars while the  solid line shows
  the  limit case $Log  (L_{K}) =  Log (\dot  E)$.  Right  panel: the
  derived IR  efficiencies, defined as the ratio  between $L_{K}$ and
  $\dot E$,  as a function of  the dipole magnetic  field $B$ inferred
  from the NS  spin down.  Timing parameters have  been taken from the
  ATNF                                                           Pulsar
  Database (http://www.atnf.csiro.au/research/pulsar/psrcat)
  for rotation-powered pulsars, and  from Woods \& Thompson (2006) for
  the magnetars. For  both 1E 1048--5937 and SGR  1806--20 we have taken
  as a reference the average $\dot P$ value. }
\label{Lir}       
\end{figure*}

\subsection{Results}

In the left  panel of Fig.~ \ref{Lir} we have  plotted the computed IR
luminosities   $L_{K}$   for   all   the   isolated   NSs   in   Table
\ref{IRsummary},  and  the  upper  limit  for PSR  J1119--6127,  as  a
function of the  NSs rotational energy loss $\dot  E$.  From this plot
we  clearly see that  the rotation-powered  pulsars and  the magnetars
cluster in quite distinct regions  of the diagram.  In particular, PSR
J1119--6127 is definitely closer  to the group of the rotation-powered
pulsars than to the magnetars  one, which would suggest its connection
with the formers rather than with the latters.

From a general point of view, it is clear that there is no substantial
difference between the IR luminosity  of the magnetars and that of the
young  ($\le  5,000$ years)  rotation-powered  pulsars  (Crab and  PSR
B1509--58), which have all luminosities $L_{K} \sim 10^{30} - 10^{32}$
erg s$^{-1}$ to be compared  with $L_{K} \sim 10^{26} - 10^{28}$ erg
s$^{-1}$ of the older rotation-powered pulsars (Vela, PSR B0656+14 and
Geminga).  For the rotation-powered pulsars Fig.~ \ref{Lir} shows that
the IR  luminosity correlates rather  well with the  rotational energy
loss, with $L_{K} \propto \dot E^{1.3 \pm 0.04}$.  This correlation is
similar to  the one found  for the optical luminosity,  i.e.  $L_{opt}
\propto \dot E^{1.6 \pm 0.2}$ (see, e.g.  Kramer 2004), which confirms
that  the IR emission  of rotation-powered  pulsars, like  the optical
one, is mostly magnetospheric.

Instead, for the magnetars the scatter of the points does not allow to
recognize a correlation between $L_{K}$ and $\dot E$.  However, if the
magnetars' IR emission were also powered by their rotational energy
they would be much more efficient IR emitters than the
rotation-powered pulsars, with IR luminosities much closer to their
$\dot E$.  In particular, we note that, if the distance lower limit of
5 kpc is confirmed, the IR luminosity of 4U 0142+61 could be
comparable to its its $\dot E$, making it the intrinsically more
luminous magnetar.  This intrinsically larger IR output could be
explained by the presence of an additional source of emission which,
at least in the case of 4U 0142+61, might be identified with a
possible fossil disk (Wang et al.  2006).  The same might be true also
for the other magnetar with the highest IR luminosity (SGR 1806--20),
while for the others the presence of a surrounding fallback disk
appears less compelling.  Alternatively, it is possible that the IR
emission of magnetars is powered, as it is in the X-rays, by the star
magnetic field rather than by its rotation.  We have plotted in the
right panel of Fig.~ \ref{Lir} the IR emission efficiency as a
function of the dipole magnetic field $B$.  Despite the scatter of the
points, it is clear that the magnetic field does imply a larger IR
emission efficiency for the magnetars than for the rotation-powered
pulsars.  We thus speculate that, although the contribution of a disk
cannot be a priori ruled out, the IR emission of the magnetars is
substantially driven by the magnetic field.  In particular, we note
that with a magnetic field $B \sim 4.1 \times 10^{13}$ G, one might
expect for PSR J1119--6127 a magnetar-like IR emission efficiency,
while it is at least one order of magnitude lower.  This makes PSR
J1119--6127, once again, more similar to the rotation-powered pulsars
than to the magnetars.  This might suggest that the actual magnetic
field of PSR J1119--6127 is lower than the measured one and that a
torque from a disk might have indeed affected the pulsar's spin
down.  However, we note that,
given the disk accretion rate compatible with the maximum
torque and the low X-ray luminosity of the pulsar (see \S3.1), the
contribution of such a disk to the total IR flux would likely be low enough for the pulsar IR emission to be dominated
by the magnetospheric component, as in the classical, rotation-powered radio pulsars.

\section{Conclusions}

We have  reported on deep IR  observations  performed  with the ESO \vlt\ to constrain
the  presence  of a  fallback  disk around the  high magnetic field
radio pulsar PSR J1119--6127.
 No IR  emission has  been detected at  the pulsar's position  down to
limiting magnitudes of $J \sim$24, $H \sim$23 and $K_s \sim$22.  These
upper limits have been compared  with the expected IR spectrum emitted
from a fallback disk, which we  have computed using the disk models of
Perna  et al.   (2000).  We  have found  that the  current  flux upper
limits only  rule out  a fallback disk  with $\dot{M}\ga  3\times 10^{16}$
g s$^{-1}$.  However, a disk with  an accretion rate of $\sim 10^{15}$ g s$^{-1}$
can still account for the rotation energy loss of the pulsar, hence we
cannot yet  confirm or  exclude that the  pulsar experiences  an extra
torque produced by a fallback disk, and that the value of the magnetic
field inferred from $P$ and  $\dot{P}$ is thus overestimated.  We have
also compared the upper limit  on the IR luminosity of PSR J1119--6127
with  the measured  IR  luminosities of  rotation-powered pulsars  and
magnetars.   While  magnetars  are  intrinsically  more  efficient  IR
emitters  than  rotation-powered pulsars,  probably  because of  their
higher  magnetic field,  we  have  found that  the  relatively low  IR
emission efficiency  of PSR J1119--6127  makes it more similar  to the
latters than  to the formers.  Although not strictly  compelling, this
might be  an indication  of a magnetic  field actually lower  than the
measured one.

\begin{acknowledgements}
RPM thanks S. Zane for her comments and suggestions. NR is supported by an NWO Post-doctoral Fellowship and a Short Term Visiting Fellowship awarded by the University of Sydney.

\end{acknowledgements}

\end{document}